\documentclass[preprint,showpacs,preprintnumbers,amsmath,amssymb]{revtex4}

\usepackage{graphicx}
\usepackage{dcolumn}
\usepackage{bm}
\begin{document}

\title[]{Magnetic-field-induced spin-crossover transition in [Mn$^{\textrm{III}}$(taa)] studied by X-ray absorption spectroscopy}

\author{Jim Long Her}
\affiliation{Institute for Solid State Physics, University of Tokyo, Kashiwa, Chiba 277-8581, Japan}
\author{Yasuhiro H. Matsuda}
\altaffiliation[Also at ]{PRESTO, Japan Science and Technology Agency, Saitama, Saitama 332-0012}
\affiliation{Institute for Solid State Physics, University of Tokyo, Kashiwa, Chiba 277-8581, Japan}
\author{Motohiro Nakano}
\affiliation{Division of Applied Chemistry, Graduate School of Engineering,
 Osaka University, Suita, Osaka 565-0871, Japan}
\author{Yasuhiro Niwa}
\affiliation{Photon Factory, Institute of Materials Structure Science, High Energy Accelerator Research Organization, Tsukuba, Ibaraki 305-0801, Japan}
\author{Yasuhiro Inada}
\altaffiliation[Present address, ]{College of Life Science, Ritsumeikan University, Kusatsu, Shiga 525-8577, Japan}
\affiliation{Photon Factory, Institute of Materials Structure Science, High Energy Accelerator Research Organization, Tsukuba, Ibaraki 305-0801, Japan}

\begin{abstract}
The X-ray absorption near-edge structure (XANES) of Mn in a spin-crossover compound, [Mn$^{\textrm{III}}$(taa)], was studied in pulsed high magnetic fields up to 37~T. By applying magnetic fields to the low-temperature low-spin (LS) state, significant changes in the spectra were observed, suggesting a magnetic-field-induced spin-crossover to the high-spin (HS) state.
At low temperatures, the magnetic field dependence of the changes in the spectra exhibited hysteresis. Furthermore, when the magnetic field was set to zero, a considerable remanent component was observed. The energy barrier of the HS $\rightarrow$ LS transition was evaluated from the temperature dependence of the decay time of the remanent signal. The energy barrier of the transition was found to be 134~K, which is notably lower than that for other spin-crossover compounds reported previously. Since the fraction of the field-induced HS state was at most 30\% and the thermodynamic macroscopic field-induced phase transition was expected to occur in fields higher than 55~T, the observed field-induced transition at low temperatures down to 17~K could be understood as a localized microscopic transition at the single-molecular level.
\end{abstract}

\maketitle

\section{Introduction}
The family of transition-metal complexes with 4$\sim$7 d-electrons usually exhibits bistability of high-spin (HS) and low-spin (LS) configurations due to competition between the Pauli exclusion principle and the crystal field energy. Some such molecules undergo a spin-crossover transition by
altering the external environment, for example, changing temperature,\cite{Gutlich1994} emitting light\cite{Herrera2004} or generating a magnetic field.\cite{Negre2001}
The presently studied compound [Mn$^{\textrm{III}}$(taa)] [H$_{3}$taa=tris(1-(2-azolyl)-2-azabuten-4-yl)amine] is known to undergo a spin-crossover transition from the HS to LS state with decreasing temperature ($T_{c}\sim$46 K), which has been confirmed by various experiments.\cite{Yann2000,Sim1981} Typically, the entropy change for the transition from the HS to LS state arises from not only spin degeneracy but also vibrational entropy. However, in the case of [Mn$^{\textrm{III}}$(taa)], another degree of freedom is also important. Dielectric measurements have shown that the existence of dynamic Jahn-Teller distortion in the HS state contributes to the change in entropy.\cite{Nakano2002,Nakano2003} These studies have also revealed a paraelectric phase in the HS state, in which electric dipoles originate from the dynamic Jahn-Teller distortion.
Moreover, it has been suggested that this compound possibly undergoes a ferroelectric transition at an expected transition temperature of $T_{FO}$ = 26 K, if no spin-crossover transition occurs.
In other words, the metastable ferroelectric ordering (FO) state is expected to become stable when the LS state is suppressed by application of another perturbation,
such as light irradiation or a magnetic field.
Although it has been reported that the magnetic field-induced spin-crossover transition (LS $\rightarrow$ HS) takes place in high fields of up to 55~T
at low temperatures just below $T_c$,\cite{Kimura2003,Kimura2005} the FO state has not been observed experimentally to date.
According to a theoretical prediction,\cite{Kimura2005} higher magnetic fields are required to observe the LS $\rightarrow$ FO transition at
low temperatures below $T_{FO}$.

Since dynamic Jahn-Teller distortion in the HS state plays an important role in the spin-crossover transition and the possible FO phase in
[Mn$^{\textrm{III}}$(taa)], an examination of the local structure on a microscopic level is highly desired.
X-ray absorption near-edge structure (XANES) spectroscopy is a powerful method for investigating such local structures.
There have been several reports on the use of XANES spectroscopy to study the local deformation of a crystal lattice due to the Jahn-Teller effect
in magnetic materials such as perovskite manganites.\cite {Bridges2000,Qian2001,Sanchez2003}
In addition, the dynamics of the light-induced excited spin state of an iron-based spin-crossover material has been studied by picosecond X-ray absorption spectroscopy and X-ray diffraction analysis, as well as by time-resolved optical spectroscopy. The mechanism for inducing the HS state and returning to the LS state was also investigated.\cite{lorenc:028301,gawelda:057401,gawelda:124520,Ch.Bressler01232009,Matthias2008} However, studies on the dynamics of field-induced spin-crossover are rare since the magnetic fields induce a much smaller heating effect on the samples in comparison with light irradiation does.

In the present study, we used XANES spectroscopy to investigate the magnetic-field-induced spin-crossover in [Mn$^{\textrm{III}}$(taa)] by observing the difference between the local structures in the HS and LS states.
We also measured the temperature dependence of the XANES in zero magnetic field to obtain typical spectra for the HS and LS phases.
When we applied a pulsed high magnetic field to the low-temperature LS state, the spectra significantly changed at a certain magnetic field, and
these changes were enhanced as the field increased.
The field-induced changes in the XANES spectra were found to be similar to that observed for the temperature-driven spin-crossover transition.
Moreover, the field dependence of the changes in the XANES spectra exhibited distorted hysteresis and a considerable remanent
component, suggesting that the relaxation time of the transition was comparable to the duration time of the magnetic field (around 1 ms).
We discuss the nature of the field-induced transitions on the basis of the magnetic field dependence of the XANES spectra at various temperatures.
The observed relaxation behavior of the field-induced spectral changes were analyzed by a simple model, and the energy barrier between the
HS and LS states was determined.

\begin{figure}
  \begin{center}
  \includegraphics[width=3.3in]{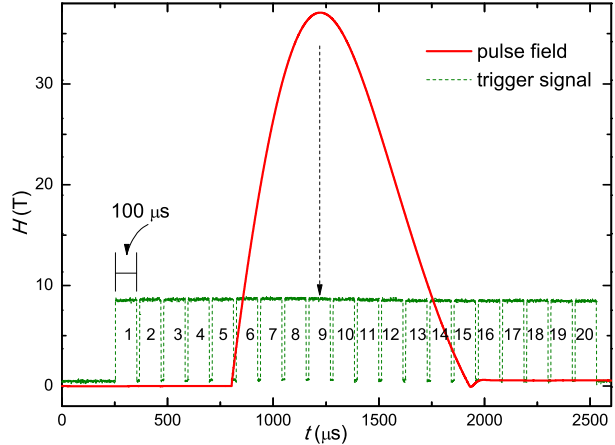}
  \end{center}
  \caption{\label{pulse40T}(Color online) Red solid curve is a representative pulsed field recorded over time. Green dashed curve shows the measurement periods of the multichannel detector.}
\end{figure}

\section{Experiment}
Polycrystalline samples were synthesized by the same method as described in the literature.\cite{Sim1981,Yann2000}
For X-ray transmission measurements, the sample was ground into a fine powder and mixed well with epoxy resin (Stycast 1266). This mixture was then compressed into a thin pellet with an effective thickness of $\sim$110~$\mu$m.
We used a small pulsed magnet installed in the NW2A beamline at the Photon Factory in Tsukuba, Japan.\cite{Ouyang2009}
A wavelength dispersive spectrometer (a so-called DXAFS spectrometer) \cite{Inada2006} and
a silicon microstrip-based X-ray detector (XSTRIP) \cite{Salvini2005} were used to obtain the absorption spectra.
A two-stage cryocooler was used to cool the sample as well as the magnet.
The sample temperature could be set between 300~K and 17~K, while the temperature of the magnet was about 90~K.
The magnet could generate pulsed fields of up to 37 T for a duration time of $\sim$1~ms.
In each discharge process, the multichannel detector recorded the spectrum twenty times discretely (twenty frames),
where the recording time window was 100~$\mu$s with a dead time of 14.4~$\mu$s between successive frames.
Using a delay trigger controller, we set the maximum field such that it always coincided with the ninth frame.
The pulsed field and recording frames of a typical discharge process are shown in Fig.~\ref{pulse40T}.
Frames 1$\sim$5 were recorded before generating the pulsed fields, and correspond to the zero-field spectrum.
We selected frame 3 as the reference spectrum for zero field.

\section{Results}
\subsection{Temperature dependence}
\begin{figure}
  \begin{center}
  \includegraphics[width=3.3in]{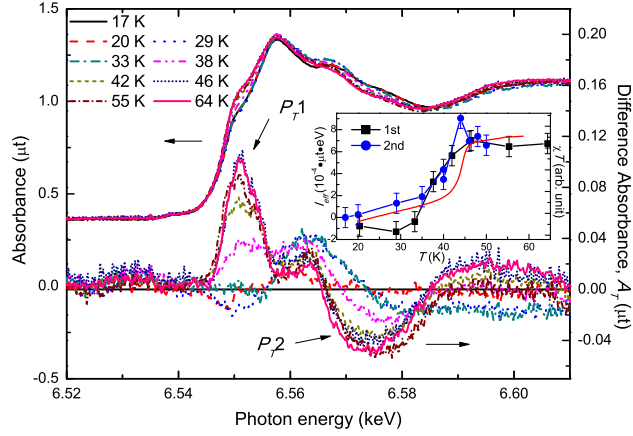}
  \end{center}
  \caption{(Color online) Temperature-dependent Mn K-edge XANES spectra (left axis) and difference spectra, $A_{T}$ (right axis), in zero field. Inset: Black line with squares and blue line with circles denote temperature-dependent $I_{eff}$ curves that were performed in different beam time; red solid line denotes temperature-dependent $\chi$ $T$ curve.}\label{DXAFSvsT}
\end{figure}

Figure \ref{DXAFSvsT} shows temperature-dependent Mn K-edge XANES spectra in zero field. The shape of the absorption edge is similar to
those in Mn$^{\texttt{III}}$ compounds with organic ligands.\cite{Hend2001}
Significant changes in the spectra were observed as the temperature increases. The difference spectra, $A_{T}$, are shown in the same figure to highlight these changes, where $A_{T}$ = Absorbance($T$) - Absorbance(17~K).
The $A_{T}$ curves show a positive peak, $P_{T}1$, at around 6.550~keV and a negative peak, $P_{T}2$, at around 6.575~keV with increasing
temperature. Interestingly, the $P_{T}1$ peak is barely observable at $T$ $\leq$ 33 K, indicating the absorbance is only
slightly different from that at 17~K. Then, the $P_{T}1$ peak becomes increasingly pronounced from 33 to 46 K and saturates
at $T$ $\geq$ 46 K.
The effective intensity of the $P_{T}1$ peak is shown in the inset of Fig. \ref{DXAFSvsT}, which was calculated using the following equation,
\begin{equation}\label{Ieff}
{I_{eff}}=\int A_{T} \textrm{d}\epsilon ~~~\epsilon: \textrm{Photon Energy}.
\end{equation}
The $P_{T}1$ peak indicates clear transition behavior that is related to the spin-crossover transition.

The magnetic susceptibility ($\chi$) of the sample used in the XANES measurements was also measured to determine the spin-crossover temperature. In the inset of Fig. \ref{DXAFSvsT}, the red solid line denotes the temperature-dependent $\chi$ $T$. The transition temperature was found to be around 44~K, which was lower than the literature value (46~K) by 2 K. Moreover, the transition appeared to be slightly broader than that reported previously.\cite{Kimura2005}
Mixing [Mn$^{\textrm{III}}$(taa)] crystals with the epoxy resin might have induced a small strain on the sample and had a small effect on the temperature-driven spin crossover transition.
It should be noted that unlike the magnetization measurement, in which a rather sharp spin-crossover transition was observed at $T_{c}$ $\sim$44 K, the XANES spectra show a broader transition boundary.
In addition, the transition temperature of XANES spectra appears to be lower than that of the magnetization measurements. We performed the temperature-dependent experiments twice in different beam time to confirm the temperature deviation. Both results were showed in the inset of Fig. \ref{DXAFSvsT}. The two curves are alike, having the transition behavior, in which the transition temperature is $\sim$4 K smaller than that of magnetization curve. Although the reason is not clear, the sample is possibly locally warmed by X-ray irradiation, which would increase the sample temperature to higher than that of the sensor.

\begin{figure}
  \begin{center}
  \includegraphics[width=3.3in]{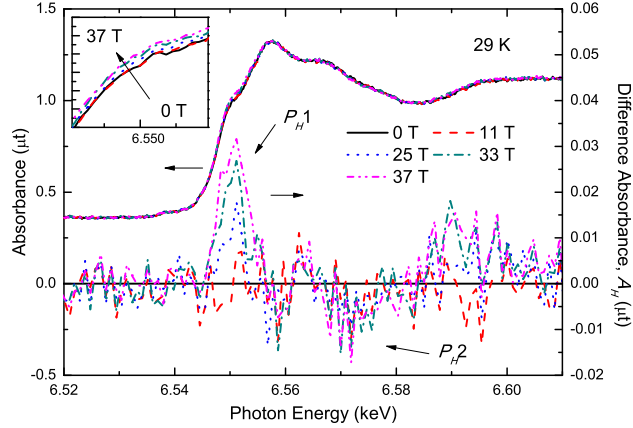}
  \end{center}
  \caption{(Color online) Field-dependent Mn K-edge XANES spectra (left axis) and the difference spectra, $A_{H}$ (right axis), at 29 K. Curves in magnetic fields of 0, 11, 25, 33, and 37 T were recorded in frames 0, 6, 7, 8, and 9, respectively. Inset shows enlarged view around 6.55 keV.}\label{DXAFSvsH40T}
\end{figure}
\begin{figure*}
  \begin{center}
  \includegraphics[width=6.5in]{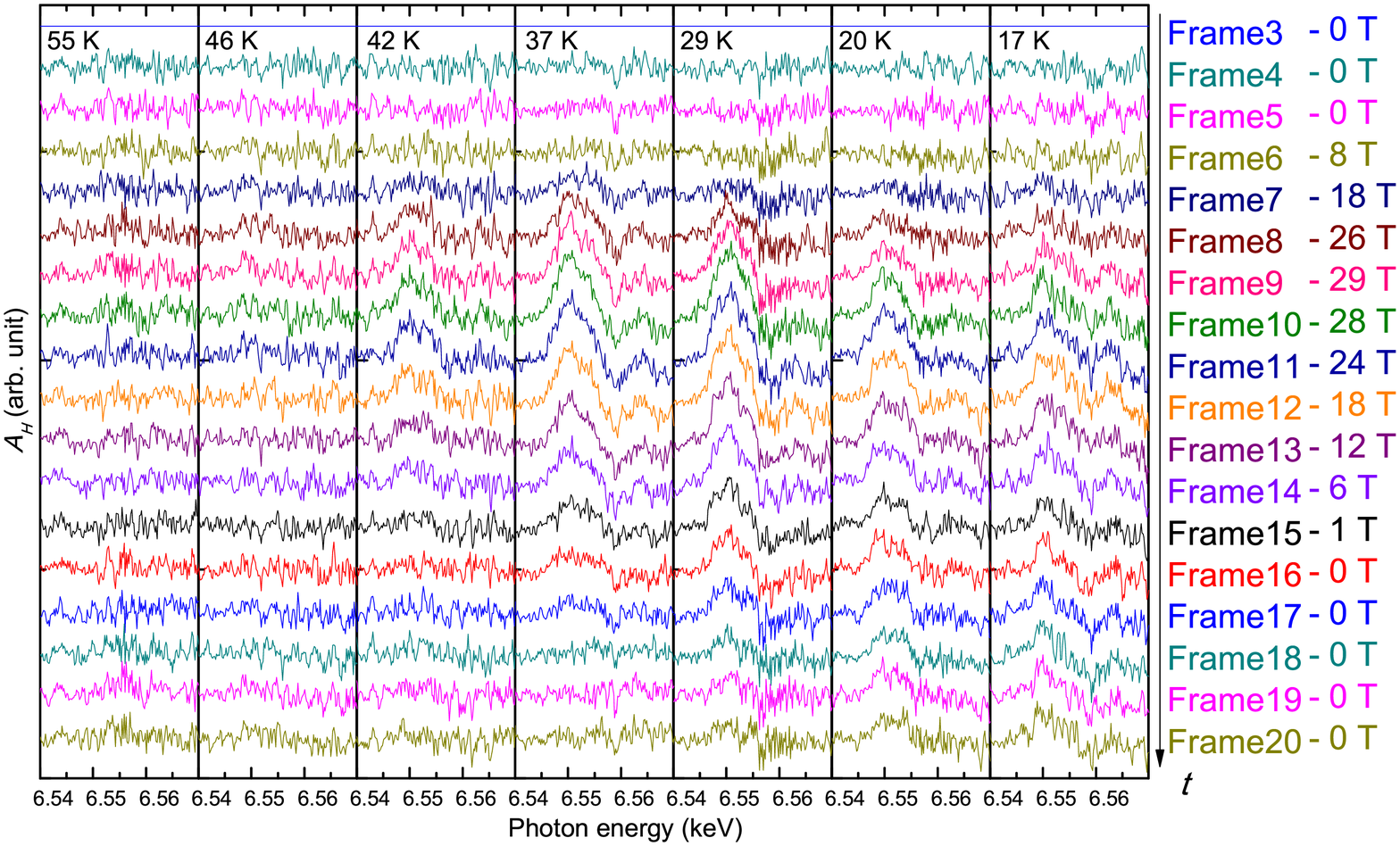}
  \end{center}
  \caption{(Color online) $A_{H}$ curves from different frames at various temperatures.}\label{Dif32T}
\end{figure*}
\subsection{Magnetic field dependence}
Field-dependent Mn K-edge XANES spectra at 29 K are shown in Fig.~\ref{DXAFSvsH40T}, in which the zero-field state is the LS state. The difference spectra, $A_{H}$, are shown in the same figure [$A_{H}$ = Absorbance($H$) - Absorbance(0 T)]. The high-field $A_{H}$ curves clearly show a field-induced peak, $P_{H}1$, at around 6.550 keV, which is at the same position as the $P_{T}1$ peak in the $A_{T}$ curves, suggesting that these peaks have the same origin. On the other hand, at around 6.575 keV, the high-field $A_{H}$ curves show an extremely weak peak, $P_{H}2$, which may correspond to the
$P_{T}2$ peak. It should be noted that, even in the largest field, the field-induced changes in the XANES spectra were small (about one-third of the thermally induced change).

Figure~\ref{Dif32T} shows the $A_{H}$ curves of the different frames at various temperatures. We only focus on the energy range around the
$P_{H}1$ peak, where several interesting phenomena can be seen. First, at $T$ $\geq$ 46 K, the $A_{H}$ curves do not show any significant field-dependent behavior. This reveals that there is no field-induced state in the temperature range of the HS state.
Second, at $T$ $\leq$ 42 K, the field-induced $P_{H}1$ peak appears. In addition, the intensity of the $P_{H}1$ peak increases with increasing magnetic field, and then decreases as the field is decreased. Third, and most importantly, the $P_{H}1$ peak remains after the pulsed field returns to zero at low temperatures (see results after frame 16).
It was also found that this remanent component decays with increasing time.
These phenomena suggest that a field-induced metastable HS state is alive at zero magnetic field and low temperatures.

\section{Discussions}
The temperature-dependent XANES spectra in zero field, as shown in Fig. \ref{DXAFSvsT},
are attributed to a spin-crossover transition.
In a previous report, dynamic Jahn-Teller distortion was suggested to exist in the HS state.\cite{Nakano2002} Therefore, the lattice structure
is slightly deformed by dynamic Jahn-Teller distortion, resulting in the changes in the XANES spectra. We take the $I_{eff}$ value in the high-temperature region as a standard value for the 100\% HS state.
In addition, it has been suggested that thermal fluctuation could excite some LS molecules to HS state even in the low-temperature LS phase.\cite{Nakano2002,Kimura2005}
These embedded HS components are expected to play an important role in the field-induced transition.

In the field-dependent measurement, the applied field of up to 37 T was not as high as that used in previous high-field magnetization studies.\cite{Kimura2003,Kimura2005}
Therefore, the fully field-induced macroscopic spin-crossover phase transition could not be reached.
However, the observation of the $P_{H}1$ and $P_{H}2$ peaks at low temperatures indicates that the local lattice structure was
distorted by magnetic fields.
One possible explanation is that in magnetic fields less than 37~T, the LS $\rightarrow$ HS transition occurs locally on a molecular level, while the macroscopic spin-crossover phase transition does not occur.
This is not unreasonable because the fraction of the field-induced HS component was at most 30\verb"%", based on the changes in the spectra
described above.

\begin{figure}
  \begin{center}
  \includegraphics[width=3.3in]{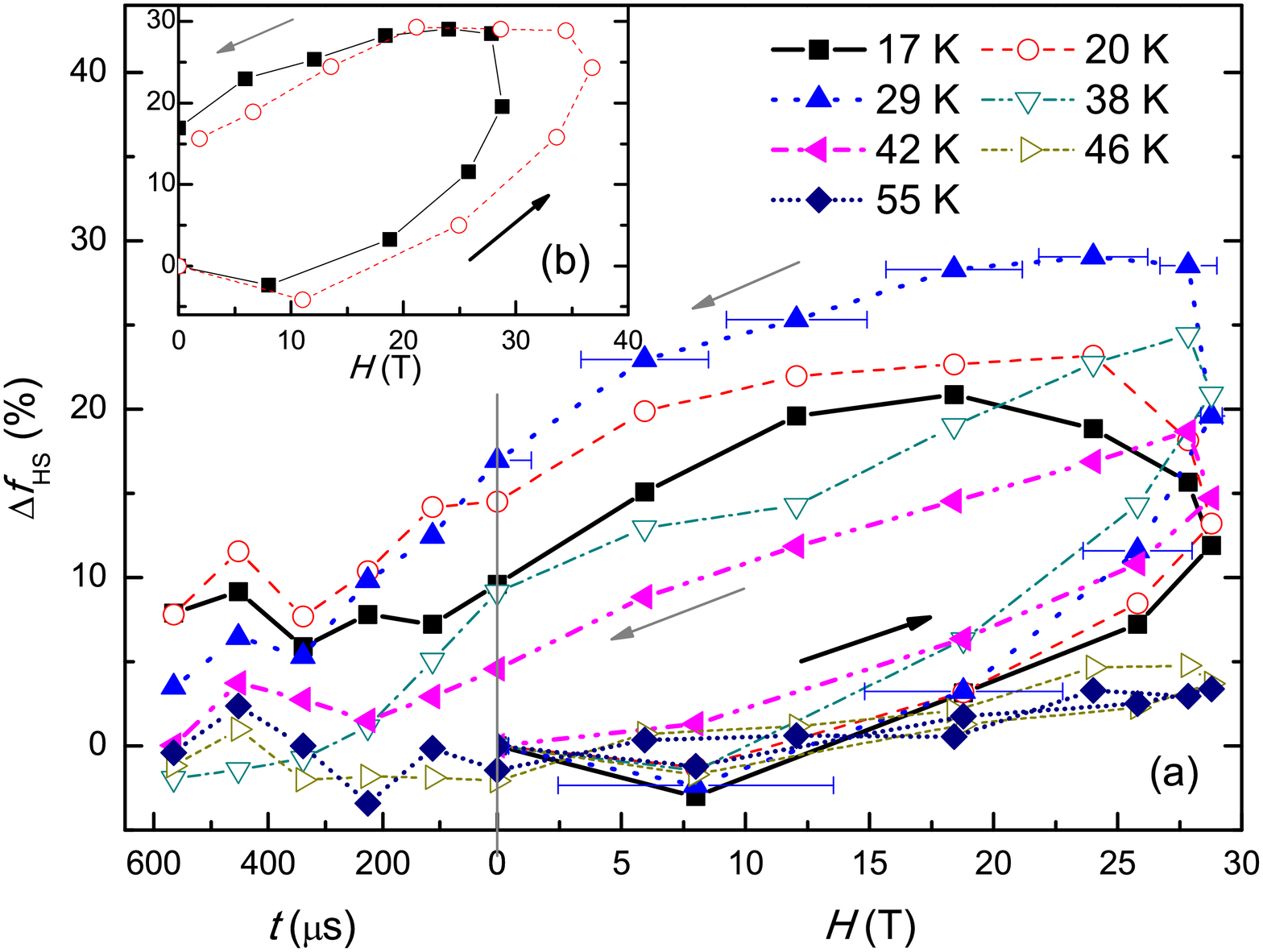}
  \end{center}
  \caption{(Color online)(a) (left panel) $\Delta \textit{f}_{\texttt{HS}}$ values of last five frames as a function of time at different temperatures. (right panel) Field-dependent $\Delta \textit{f}_{\texttt{HS}}$ at different temperatures under maximum field of 29 T. (b) $\Delta \textit{f}_{\texttt{HS}}$ curves at 29 K under different maximum fields (29 and 37 T). Bold and thin arrows indicate increasing field and decreasing field, respectively.}\label{IvsH}
\end{figure}

In order to better compare the field-induced changes in the XANES spectra, we calculated the effective intensity of the $A_{H}$ spectra by using a same equation as Eq. \ref{Ieff}. Only the energy range of the $P_{H}1$ peak was taken into account. The calculated $I_{eff}$ values can be considered as the amount of the induced HS component. As we mentioned in the previous paragraph, the $I_{eff}$ value at high temperature
was taken as the 100\verb"%" HS state; therefore, we can obtain the fraction of the field-induced HS state ($\Delta \textit{f}_{\texttt{HS}}$) from the $I_{eff}$ values. The field-dependent $\Delta \textit{f}_{\texttt{HS}}$ curves at different temperatures are shown in
the right panel of Fig.~\ref{IvsH}(a).
Clearly, the high temperature curves (46 and 55 K) show no field dependence (no hysteresis).
On the other hand, the $\Delta \textit{f}_{\texttt{HS}}$ curves exhibit distorted hysteresis at low temperatures. It is noteworthy that the largest $\Delta \textit{f}_{\texttt{HS}}$ was $\sim$30\verb"%" at 29 K and 24 T.
Even at 17 K, the $\Delta \textit{f}_{\texttt{HS}}$ value was $\sim$20\verb"%" in high fields. The $I_{eff}$ curves for different maximum pulse fields at 29 K are shown in Fig.~\ref{IvsH}.
Both curves show similar distorted hysteresis loops.

The fraction of the induced HS state at low temperature was unusually large in the observed sample
in comparison with other compounds such as [Fe(phen)$_{2}$(NCS)$_{2}$] and [Co(H$_{2}$(fsa)$_{2}$en)py$_{2}$].\cite{Bousseksou2000,Bousseksou2002,Bousseksou2004} Those compounds also undergo magnetic-field-induced LS $\rightarrow$ HS transitions; however, the fraction of the induced HS state is almost zero when the temperature is far below $T_{c}$.

\begin{figure}
  \begin{center}
  \includegraphics[width=3.3in]{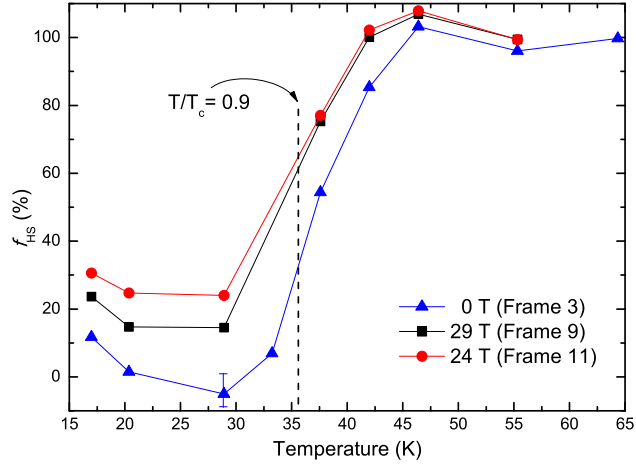}
  \end{center}
  \caption{(Color online)Temperature dependence of the fraction of HS state in 0, 29 and 24 T.}\label{fHSvsT}
\end{figure}

In Fig.~\ref{fHSvsT}, we showed the temperature dependence of the fraction of HS state ($\textit{f}_{\texttt{HS}}$) in selected magnetic fields, according to the field-dependent $\Delta \textit{f}_{\texttt{HS}}$ curves. This figure clearly shows that the field-induced HS state is significant even in low temperature region. In previous reports, Bousseksou $et~al.$ observe the production of the HS state at temperatures only near $T_c$ ($\frac{T}{T_{c}}~>~0.90$),\cite{Bousseksou2000,Bousseksou2002,Bousseksou2004} while we observed a relatively large 20\verb"%" field-induced HS state at 17 K, which was much lower than $T_c\sim$44~K ($\frac{T}{T_{c}}~\sim~0.4$). In addition, this behavior is also different with the field dependence of the $\chi$T-curve observed in the present compound,\cite{Kimura2005} where the field-induced change only existences in $\frac{T}{T_{c}}~>~0.95$.
These observations at low temperatures are unique and not well explained by the macroscopic phase transition model.

Two types of relaxation behavior can be seen in
the right and left panels of Fig. \ref{IvsH} (a).
First, in the right panel,
the maximum $\Delta \textit{f}_{\texttt{HS}}$ values do not coincide with the maximum field, showing a delay of $\sim$200 $\mu$s. According to the literature,\cite{Bousseksou2000,Negre2000,Negre2001,Bousseksou2002,Bousseksou2004}
similar phenomena have been observed for other compounds that undergo a  spin-crossover transition (LS $\rightarrow$ HS) in a pulsed field at temperatures slightly lower than the spin-crossover transition temperature.
Bousseksou $et~al.$ has suggested that the magnetic triggering transition is a macroscopic cooperative characteristic of the system.\cite{Bousseksou2000} The intermolecular interaction and vibronic coupling, as well as the energy gap between HS and LS states, are important and strongly affect the fraction of the induced HS state and the kinetic delay time. Two typical samples, [Fe(phen)$_{2}$(NCS)$_{2}$] and [Co(H$_{2}$(fsa)$_{2}$en)py$_{2}$], have been investigated, the energy gaps (triggered HS states) of which were found to be 1050 K (15\verb"%")
and 620 K (60\verb"%"),
respectively.\cite{Bousseksou2002,Bousseksou2004}
For our sample, the effective HS-LS energy gap was 340 K,\cite{Nakano2002} and the model suggests that a large fraction of the induced HS state should be observed around $T_{c}$. However, the largest fraction of the field-induced HS state in our experiment was only 30\verb"%" at 29 K.


\begin{figure}
  \begin{center}
  \includegraphics[width=3.3in]{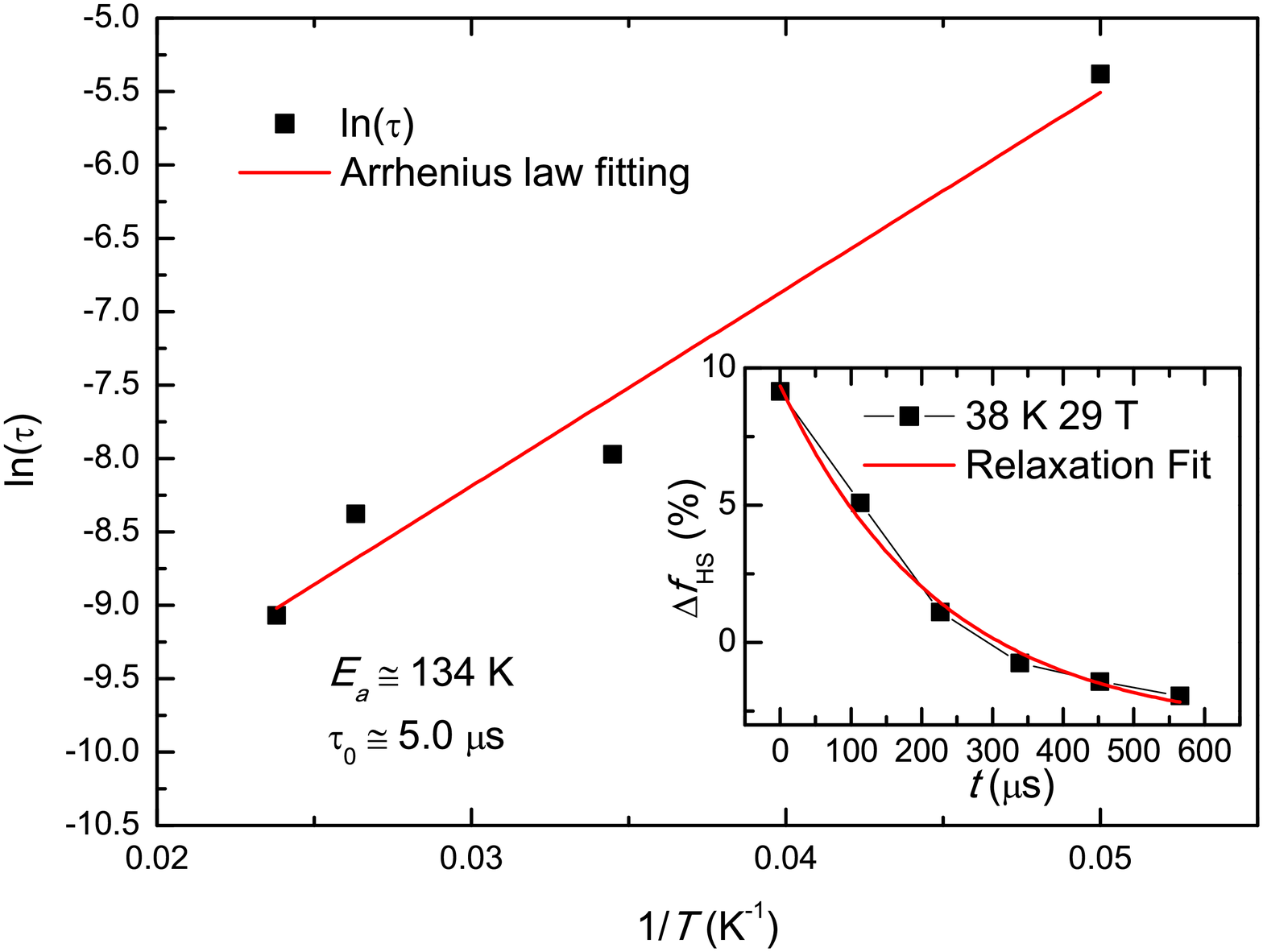}
  \end{center}
  \caption{(Color online) ln($\tau$) vs $T^{-1}$ plot. Red solid line denotes the Arrhenius law fit. Inset: Plot of $\Delta \textit{f}_{\texttt{HS}}$ as function of $t$. Red solid line is the exponential relaxation fit.}\label{ALawFit32T}
\end{figure}
A second noteworthy relaxation behavior occurs when the magnetic field decreases to zero (see left panel of Fig. \ref{IvsH}).
The value of $\Delta \textit{f}_{\texttt{HS}}$ does not immediately become zero.
A finite value of remanent $\Delta \textit{f}_{\texttt{HS}}$ exists and decays with time. The time-dependent relaxation can be clearly seen in the curves for 20, 29, and 38 K. In order to better understand this phenomenon, we further analyzed the relaxation behavior with a typical relaxation exponential equation:
\begin{equation}\label{relaxation}
    \Delta \textit{f}_{\texttt{HS}}=\Delta \textit{f}_{\texttt{HS}}^{~0}+\Delta \textit{f}_{\texttt{HS}}^{~1}\exp(\frac{-t}{\tau}),
\end{equation}
where $\tau$ is the relaxation characteristic time. The inset of Fig. \ref{ALawFit32T} shows the time-dependent $\Delta \textit{f}_{\texttt{HS}}$ curve for 38 K and the fitting result, which agrees well with the experimental results.
The fit with the relaxation exponential equation agrees well with the data for 20 $\sim$42 K. However, at 17 K, the time-dependent $\Delta \textit{f}_{\texttt{HS}}$ is almost constant, suggesting an extremely large $\tau$, which cannot be fitted well. The temperature dependence of $\tau$ is plotted in Fig. \ref{ALawFit32T}.

The HS $\rightarrow$ LS relaxation behavior has been studied for many compounds.\cite{Hauser2004}
Classical Arrhenius-type activation is the simplest model of the relaxation process, which is usually applied to a two-state excitation. As mentioned above, the field-induced HS state is local. Therefore, we suggest the relaxation behavior originates from a single-ion HS $\rightarrow$ LS relaxation process with a finite energy barrier.
Then, the temperature-dependent $\tau$ could be fitted with the Arrhenius equation:
\begin{equation}\label{ALaw}
    \ln(\tau)=\ln(\tau_{0})+\frac{E_{a}}{k_{B}T},
\end{equation}
where $\tau_{0}$ is the effective characteristic time and $E_{a}$ is the activation energy. The fitted $\frac{E_{a}}{k_{B}}$ is about 134 K and $\tau_{0}$ is about 5 $\mu$s. $E_{a}$ can be considered to be the energy barrier for the HS $\rightarrow$ LS transition. Interestingly, the obtained $E_{a}$ value is much smaller than
those reported for other compounds, which have $E_{a}$ values on the order of several hundred Kelvin.\cite{Hauser2004}
Such a low activation energy and small time constant suggest that the relaxation is likely the result of a thermal process, not a tunneling process.
Very recently, time-resolved X-ray diffraction was used to study the photoswitching dynamics of a molecular Fe(III) spin-crossover material.\cite{lorenc:028301} In this report, the relaxation time of the recovery process was on the order of milliseconds, which is close to the value of the HS $\rightarrow$ LS relaxation observed in the present study.

In the inset of Fig. \ref{ALawFit32T}, the $\Delta \textit{f}_{\texttt{HS}}$ curve has a negative value when $t~>~300~\mu$s. This feature possibly originates
from the cooling effect of adiabatic demagnetization. However, these negative $\Delta \textit{f}_{\texttt{HS}}$ values are as small as the resolution in the present experiment. In addition, in the $T$ $\leq$ $T_{c}$ region, only the curve at 38 K shows such behavior. Therefore, further studies are required to clarify the details of this phenomenon.


\section{Summary}
\label{Sum}
We carried out high-magnetic-field XANES analysis of [Mn$^{\textrm{III}}$(taa)].
The spectra showed significant differences between the high-temperature HS and low-temperature
LS states.
In field-dependent studies, the spectra also showed significant differences in high fields, even at very low
temperatures ($\frac{T}{T_{c}}~\sim~0.4$). In the previous works, the field-induced spin-crossover transitions were observed
at temperatures very close to $T_{c}$ ($\frac{T}{T_{c}}~>~0.90$).\cite{Bousseksou2000,Bousseksou2002,Bousseksou2004} Hence, our observation was a rare phenomenon. Possibly, applying magnetic fields induced the microscopic LS $\rightarrow$ HS spin-crossover transition in the molecular level. The nucleation of HS state can be initiated with a help of the thermal fluctuation. Furthermore, the field-induced changes
in the XANES spectra exhibited relaxation behavior when the magnetic field decreased to zero. We analyzed the
decay process of the remanent component
by using Arrhenius law and found that the energy barrier of the HS $\rightarrow$ LS relaxation process was
134~K.
This value is remarkably small in comparison with the energy barriers reported for other spin-crossover complexes, which are on the order of several hundred Kelvin.
We proposed that the relaxation process of [Mn$^{\textrm{III}}$(taa)] observed in the present work was
within the thermally activated region.

\bibliography{Mntaa_new}

\begin{thebibliography}{26}
\expandafter\ifx\csname natexlab\endcsname\relax\def\natexlab#1{#1}\fi
\expandafter\ifx\csname bibnamefont\endcsname\relax
  \def\bibnamefont#1{#1}\fi
\expandafter\ifx\csname bibfnamefont\endcsname\relax
  \def\bibfnamefont#1{#1}\fi
\expandafter\ifx\csname citenamefont\endcsname\relax
  \def\citenamefont#1{#1}\fi
\expandafter\ifx\csname url\endcsname\relax
  \def\url#1{\texttt{#1}}\fi
\expandafter\ifx\csname urlprefix\endcsname\relax\def\urlprefix{URL }\fi
\providecommand{\bibinfo}[2]{#2}
\providecommand{\eprint}[2][]{\url{#2}}

\bibitem[{\citenamefont{G\"{u}tlich et~al.}(1994)\citenamefont{G\"{u}tlich,
  Hauser, and Spiering}}]{Gutlich1994}
\bibinfo{author}{\bibfnamefont{P.}~\bibnamefont{G\"{u}tlich}},
  \bibinfo{author}{\bibfnamefont{A.}~\bibnamefont{Hauser}}, \bibnamefont{and}
  \bibinfo{author}{\bibfnamefont{H.}~\bibnamefont{Spiering}},
  \bibinfo{journal}{Angew. Chem. Int. Ed.} \textbf{\bibinfo{volume}{33}},
  \bibinfo{pages}{2024} (\bibinfo{year}{1994}).

\bibitem[{\citenamefont{Herrera et~al.}(2004)\citenamefont{Herrera, Marvaud,
  Verdaguer, Marrot, Kalisz, and Mathoniere}}]{Herrera2004}
\bibinfo{author}{\bibfnamefont{J.~M.} \bibnamefont{Herrera}},
  \bibinfo{author}{\bibfnamefont{V.}~\bibnamefont{Marvaud}},
  \bibinfo{author}{\bibfnamefont{M.}~\bibnamefont{Verdaguer}},
  \bibinfo{author}{\bibfnamefont{J.}~\bibnamefont{Marrot}},
  \bibinfo{author}{\bibfnamefont{M.}~\bibnamefont{Kalisz}}, \bibnamefont{and}
  \bibinfo{author}{\bibfnamefont{C.}~\bibnamefont{Mathoniere}},
  \bibinfo{journal}{Angew. Chem. Int. Ed.} \textbf{\bibinfo{volume}{43}},
  \bibinfo{pages}{5468} (\bibinfo{year}{2004}).

\bibitem[{\citenamefont{N\`{e}gre et~al.}(2001)\citenamefont{N\`{e}gre,
  Cons\'{e}jo, Goiran, Bousseksou, Varret, Tuchagues, Barbaste, Ask\'{e}nazy,
  and Haasnoot}}]{Negre2001}
\bibinfo{author}{\bibfnamefont{N.}~\bibnamefont{N\`{e}gre}},
  \bibinfo{author}{\bibfnamefont{C.}~\bibnamefont{Cons\'{e}jo}},
  \bibinfo{author}{\bibfnamefont{M.}~\bibnamefont{Goiran}},
  \bibinfo{author}{\bibfnamefont{A.}~\bibnamefont{Bousseksou}},
  \bibinfo{author}{\bibfnamefont{F.}~\bibnamefont{Varret}},
  \bibinfo{author}{\bibfnamefont{J.~P.} \bibnamefont{Tuchagues}},
  \bibinfo{author}{\bibfnamefont{R.}~\bibnamefont{Barbaste}},
  \bibinfo{author}{\bibfnamefont{S.}~\bibnamefont{Ask\'{e}nazy}},
  \bibnamefont{and} \bibinfo{author}{\bibfnamefont{J.~G.}
  \bibnamefont{Haasnoot}}, \bibinfo{journal}{Physica B}
  \textbf{\bibinfo{volume}{294-295}}, \bibinfo{pages}{91}
  (\bibinfo{year}{2001}).

\bibitem[{\citenamefont{Garcia et~al.}(2000)\citenamefont{Garcia, Kahn, Ader,
  Buzdin, Meurdesoif, and Guillot}}]{Yann2000}
\bibinfo{author}{\bibfnamefont{Y.}~\bibnamefont{Garcia}},
  \bibinfo{author}{\bibfnamefont{O.}~\bibnamefont{Kahn}},
  \bibinfo{author}{\bibfnamefont{J.-P.} \bibnamefont{Ader}},
  \bibinfo{author}{\bibfnamefont{A.}~\bibnamefont{Buzdin}},
  \bibinfo{author}{\bibfnamefont{Y.}~\bibnamefont{Meurdesoif}},
  \bibnamefont{and} \bibinfo{author}{\bibfnamefont{M.}~\bibnamefont{Guillot}},
  \bibinfo{journal}{Phys. Lett. A} \textbf{\bibinfo{volume}{271}},
  \bibinfo{pages}{145} (\bibinfo{year}{2000}).

\bibitem[{\citenamefont{Sim and Sinn}(1981)}]{Sim1981}
\bibinfo{author}{\bibfnamefont{P.~G.} \bibnamefont{Sim}} \bibnamefont{and}
  \bibinfo{author}{\bibfnamefont{E.}~\bibnamefont{Sinn}}, \bibinfo{journal}{J.
  Am. Chem. Soc.} \textbf{\bibinfo{volume}{103}}, \bibinfo{pages}{241}
  (\bibinfo{year}{1981}).

\bibitem[{\citenamefont{Nakano et~al.}(2002)\citenamefont{Nakano, Matsubayashi,
  and Matsuo}}]{Nakano2002}
\bibinfo{author}{\bibfnamefont{M.}~\bibnamefont{Nakano}},
  \bibinfo{author}{\bibfnamefont{G.~E.} \bibnamefont{Matsubayashi}},
  \bibnamefont{and} \bibinfo{author}{\bibfnamefont{T.}~\bibnamefont{Matsuo}},
  \bibinfo{journal}{{Phys. Rev. B}} \textbf{\bibinfo{volume}{66}},
  \bibinfo{pages}{{212412}} (\bibinfo{year}{2002}).

\bibitem[{\citenamefont{Nakano et~al.}(2003)\citenamefont{Nakano, Matsubayashi,
  and Matsuo}}]{Nakano2003}
\bibinfo{author}{\bibfnamefont{M.}~\bibnamefont{Nakano}},
  \bibinfo{author}{\bibfnamefont{G.~E.} \bibnamefont{Matsubayashi}},
  \bibnamefont{and} \bibinfo{author}{\bibfnamefont{T.}~\bibnamefont{Matsuo}},
  \bibinfo{journal}{Adv. Quantum Chem.} \textbf{\bibinfo{volume}{44}},
  \bibinfo{pages}{617} (\bibinfo{year}{2003}).

\bibitem[{\citenamefont{Kimura et~al.}(2003)\citenamefont{Kimura, Otani,
  Narumi, Kindo, Nakano, and Matsubayashi}}]{Kimura2003}
\bibinfo{author}{\bibfnamefont{S.}~\bibnamefont{Kimura}},
  \bibinfo{author}{\bibfnamefont{T.}~\bibnamefont{Otani}},
  \bibinfo{author}{\bibfnamefont{Y.}~\bibnamefont{Narumi}},
  \bibinfo{author}{\bibfnamefont{K.}~\bibnamefont{Kindo}},
  \bibinfo{author}{\bibfnamefont{M.}~\bibnamefont{Nakano}}, \bibnamefont{and}
  \bibinfo{author}{\bibfnamefont{G.}~\bibnamefont{Matsubayashi}},
  \bibinfo{journal}{J. Phys. Soc. Jpan.} \textbf{\bibinfo{volume}{72SB}},
  \bibinfo{pages}{122} (\bibinfo{year}{2003}).

\bibitem[{\citenamefont{Kimura et~al.}(2005)\citenamefont{Kimura, Narumi,
  Kindo, Nakano, and Matsubayashi}}]{Kimura2005}
\bibinfo{author}{\bibfnamefont{S.}~\bibnamefont{Kimura}},
  \bibinfo{author}{\bibfnamefont{Y.}~\bibnamefont{Narumi}},
  \bibinfo{author}{\bibfnamefont{K.}~\bibnamefont{Kindo}},
  \bibinfo{author}{\bibfnamefont{M.}~\bibnamefont{Nakano}}, \bibnamefont{and}
  \bibinfo{author}{\bibfnamefont{G.-e.} \bibnamefont{Matsubayashi}},
  \bibinfo{journal}{Phys. Rev. B} \textbf{\bibinfo{volume}{72}},
  \bibinfo{pages}{064448} (\bibinfo{year}{2005}).

\bibitem[{\citenamefont{Bridges et~al.}(2000)\citenamefont{Bridges, Booth,
  Kwei, Neumeier, and Sawatzky}}]{Bridges2000}
\bibinfo{author}{\bibfnamefont{F.}~\bibnamefont{Bridges}},
  \bibinfo{author}{\bibfnamefont{C.~H.} \bibnamefont{Booth}},
  \bibinfo{author}{\bibfnamefont{G.~H.} \bibnamefont{Kwei}},
  \bibinfo{author}{\bibfnamefont{J.~J.} \bibnamefont{Neumeier}},
  \bibnamefont{and} \bibinfo{author}{\bibfnamefont{G.~A.}
  \bibnamefont{Sawatzky}}, \bibinfo{journal}{Phys. Rev. B}
  \textbf{\bibinfo{volume}{61}}, \bibinfo{pages}{R9237} (\bibinfo{year}{2000}).

\bibitem[{\citenamefont{Qian et~al.}(2001)\citenamefont{Qian, Tyson, Kao,
  Croft, Cheong, Popov, and Greenblatt}}]{Qian2001}
\bibinfo{author}{\bibfnamefont{Q.}~\bibnamefont{Qian}},
  \bibinfo{author}{\bibfnamefont{T.~A.} \bibnamefont{Tyson}},
  \bibinfo{author}{\bibfnamefont{C.-C.} \bibnamefont{Kao}},
  \bibinfo{author}{\bibfnamefont{M.}~\bibnamefont{Croft}},
  \bibinfo{author}{\bibfnamefont{S.-W.} \bibnamefont{Cheong}},
  \bibinfo{author}{\bibfnamefont{G.}~\bibnamefont{Popov}}, \bibnamefont{and}
  \bibinfo{author}{\bibfnamefont{M.}~\bibnamefont{Greenblatt}},
  \bibinfo{journal}{Phys. Rev. B} \textbf{\bibinfo{volume}{64}},
  \bibinfo{pages}{024430} (\bibinfo{year}{2001}).

\bibitem[{\citenamefont{Sanchez et~al.}(2003)\citenamefont{Sanchez, Subias,
  Garcia, and Blasco}}]{Sanchez2003}
\bibinfo{author}{\bibfnamefont{M.~C.} \bibnamefont{Sanchez}},
  \bibinfo{author}{\bibfnamefont{G.}~\bibnamefont{Subias}},
  \bibinfo{author}{\bibfnamefont{J.}~\bibnamefont{Garcia}}, \bibnamefont{and}
  \bibinfo{author}{\bibfnamefont{J.}~\bibnamefont{Blasco}},
  \bibinfo{journal}{Phys. Rev. Lett.} \textbf{\bibinfo{volume}{90}},
  \bibinfo{pages}{045503} (\bibinfo{year}{2003}).

\bibitem[{\citenamefont{Lorenc et~al.}(2009)\citenamefont{Lorenc, H\'{e}bert,
  Moisan, Trzop, Servol, Cointe, Cailleau, Boillot, Pontecorvo, Wulff
  et~al.}}]{lorenc:028301}
\bibinfo{author}{\bibfnamefont{M.}~\bibnamefont{Lorenc}},
  \bibinfo{author}{\bibfnamefont{J.}~\bibnamefont{H\'{e}bert}},
  \bibinfo{author}{\bibfnamefont{N.}~\bibnamefont{Moisan}},
  \bibinfo{author}{\bibfnamefont{E.}~\bibnamefont{Trzop}},
  \bibinfo{author}{\bibfnamefont{M.}~\bibnamefont{Servol}},
  \bibinfo{author}{\bibfnamefont{M.~B.-L.} \bibnamefont{Cointe}},
  \bibinfo{author}{\bibfnamefont{H.}~\bibnamefont{Cailleau}},
  \bibinfo{author}{\bibfnamefont{M.~L.} \bibnamefont{Boillot}},
  \bibinfo{author}{\bibfnamefont{E.}~\bibnamefont{Pontecorvo}},
  \bibinfo{author}{\bibfnamefont{M.}~\bibnamefont{Wulff}},
  \bibnamefont{et~al.}, \bibinfo{journal}{Phys. Rev. Lett.}
  \textbf{\bibinfo{volume}{103}}, \bibinfo{eid}{028301} (\bibinfo{year}{2009}).

\bibitem[{\citenamefont{Gawelda et~al.}(2007)\citenamefont{Gawelda, Pham,
  Benfatto, Zaushitsyn, Kaiser, Grolimund, Johnson, Abela, Hauser, Bressler
  et~al.}}]{gawelda:057401}
\bibinfo{author}{\bibfnamefont{W.}~\bibnamefont{Gawelda}},
  \bibinfo{author}{\bibfnamefont{V.-T.} \bibnamefont{Pham}},
  \bibinfo{author}{\bibfnamefont{M.}~\bibnamefont{Benfatto}},
  \bibinfo{author}{\bibfnamefont{Y.}~\bibnamefont{Zaushitsyn}},
  \bibinfo{author}{\bibfnamefont{M.}~\bibnamefont{Kaiser}},
  \bibinfo{author}{\bibfnamefont{D.}~\bibnamefont{Grolimund}},
  \bibinfo{author}{\bibfnamefont{S.~L.} \bibnamefont{Johnson}},
  \bibinfo{author}{\bibfnamefont{R.}~\bibnamefont{Abela}},
  \bibinfo{author}{\bibfnamefont{A.}~\bibnamefont{Hauser}},
  \bibinfo{author}{\bibfnamefont{C.}~\bibnamefont{Bressler}},
  \bibnamefont{et~al.}, \bibinfo{journal}{Phys. Rev. Lett.}
  \textbf{\bibinfo{volume}{98}}, \bibinfo{eid}{057401} (\bibinfo{year}{2007}).

\bibitem[{\citenamefont{Gawelda et~al.}(2009)\citenamefont{Gawelda, Pham,
  van~der Veen, Grolimund, Abela, Chergui, and Bressler}}]{gawelda:124520}
\bibinfo{author}{\bibfnamefont{W.}~\bibnamefont{Gawelda}},
  \bibinfo{author}{\bibfnamefont{V.-T.} \bibnamefont{Pham}},
  \bibinfo{author}{\bibfnamefont{R.~M.} \bibnamefont{van~der Veen}},
  \bibinfo{author}{\bibfnamefont{D.}~\bibnamefont{Grolimund}},
  \bibinfo{author}{\bibfnamefont{R.}~\bibnamefont{Abela}},
  \bibinfo{author}{\bibfnamefont{M.}~\bibnamefont{Chergui}}, \bibnamefont{and}
  \bibinfo{author}{\bibfnamefont{C.}~\bibnamefont{Bressler}},
  \bibinfo{journal}{J. Chem. Phys.} \textbf{\bibinfo{volume}{130}},
  \bibinfo{pages}{124520} (\bibinfo{year}{2009}).

\bibitem[{\citenamefont{Bressler et~al.}(2009)\citenamefont{Bressler, Milne,
  Pham, ElNahhas, van~der Veen, Gawelda, Johnson, Beaud, Grolimund, Kaiser
  et~al.}}]{Ch.Bressler01232009}
\bibinfo{author}{\bibfnamefont{C.}~\bibnamefont{Bressler}},
  \bibinfo{author}{\bibfnamefont{C.}~\bibnamefont{Milne}},
  \bibinfo{author}{\bibfnamefont{V.-T.} \bibnamefont{Pham}},
  \bibinfo{author}{\bibfnamefont{A.}~\bibnamefont{ElNahhas}},
  \bibinfo{author}{\bibfnamefont{R.~M.} \bibnamefont{van~der Veen}},
  \bibinfo{author}{\bibfnamefont{W.}~\bibnamefont{Gawelda}},
  \bibinfo{author}{\bibfnamefont{S.}~\bibnamefont{Johnson}},
  \bibinfo{author}{\bibfnamefont{P.}~\bibnamefont{Beaud}},
  \bibinfo{author}{\bibfnamefont{D.}~\bibnamefont{Grolimund}},
  \bibinfo{author}{\bibfnamefont{M.}~\bibnamefont{Kaiser}},
  \bibnamefont{et~al.}, \bibinfo{journal}{Science}
  \textbf{\bibinfo{volume}{323}}, \bibinfo{pages}{489} (\bibinfo{year}{2009}).

\bibitem[{\citenamefont{Wolf et~al.}(2008)\citenamefont{Wolf, Gro{\ss},
  Schumann, Wolny, Sch\"{u}nemann, D{\o}ssing, Paulsen, McGarvey, and
  Diller}}]{Matthias2008}
\bibinfo{author}{\bibfnamefont{M.~M.~N.} \bibnamefont{Wolf}},
  \bibinfo{author}{\bibfnamefont{R.}~\bibnamefont{Gro{\ss}}},
  \bibinfo{author}{\bibfnamefont{C.}~\bibnamefont{Schumann}},
  \bibinfo{author}{\bibfnamefont{J.~A.} \bibnamefont{Wolny}},
  \bibinfo{author}{\bibfnamefont{V.}~\bibnamefont{Sch\"{u}nemann}},
  \bibinfo{author}{\bibfnamefont{A.}~\bibnamefont{D{\o}ssing}},
  \bibinfo{author}{\bibfnamefont{H.}~\bibnamefont{Paulsen}},
  \bibinfo{author}{\bibfnamefont{J.~J.} \bibnamefont{McGarvey}},
  \bibnamefont{and} \bibinfo{author}{\bibfnamefont{R.}~\bibnamefont{Diller}},
  \bibinfo{journal}{Phys. Chem. Chem. Phys.} \textbf{\bibinfo{volume}{10}},
  \bibinfo{eid}{057401} (\bibinfo{year}{2008}).

\bibitem[{\citenamefont{Ouyang et~al.}(2009)\citenamefont{Ouyang, Matsuda,
  Nojiri, Inada, Niwa, and Arima}}]{Ouyang2009}
\bibinfo{author}{\bibfnamefont{Z.~W.} \bibnamefont{Ouyang}},
  \bibinfo{author}{\bibfnamefont{Y.~H.} \bibnamefont{Matsuda}},
  \bibinfo{author}{\bibfnamefont{H.}~\bibnamefont{Nojiri}},
  \bibinfo{author}{\bibfnamefont{Y.}~\bibnamefont{Inada}},
  \bibinfo{author}{\bibfnamefont{Y.}~\bibnamefont{Niwa}}, \bibnamefont{and}
  \bibinfo{author}{\bibfnamefont{T.}~\bibnamefont{Arima}}, \bibinfo{journal}{J.
  Phys. :Condens. Matter} \textbf{\bibinfo{volume}{21}},
  \bibinfo{pages}{016006} (\bibinfo{year}{2009}).

\bibitem[{\citenamefont{Inada et~al.}(2007)\citenamefont{Inada, Suzuki, Niwa,
  and Nomura}}]{Inada2006}
\bibinfo{author}{\bibfnamefont{Y.}~\bibnamefont{Inada}},
  \bibinfo{author}{\bibfnamefont{A.}~\bibnamefont{Suzuki}},
  \bibinfo{author}{\bibfnamefont{Y.}~\bibnamefont{Niwa}}, \bibnamefont{and}
  \bibinfo{author}{\bibfnamefont{M.}~\bibnamefont{Nomura}},
  \bibinfo{journal}{AIP Conf. Proc.} \textbf{\bibinfo{volume}{879}},
  \bibinfo{pages}{1230} (\bibinfo{year}{2007}).

\bibitem[{\citenamefont{Salvini et~al.}(2005)\citenamefont{Salvini, Headspith,
  Thomas, Derbyshire, Dent, Rayment, Evans, Farrow, Diaz-Moreno, and
  Ponchut}}]{Salvini2005}
\bibinfo{author}{\bibfnamefont{G.}~\bibnamefont{Salvini}},
  \bibinfo{author}{\bibfnamefont{J.}~\bibnamefont{Headspith}},
  \bibinfo{author}{\bibfnamefont{S.~L.} \bibnamefont{Thomas}},
  \bibinfo{author}{\bibfnamefont{G.}~\bibnamefont{Derbyshire}},
  \bibinfo{author}{\bibfnamefont{A.}~\bibnamefont{Dent}},
  \bibinfo{author}{\bibfnamefont{T.}~\bibnamefont{Rayment}},
  \bibinfo{author}{\bibfnamefont{J.}~\bibnamefont{Evans}},
  \bibinfo{author}{\bibfnamefont{R.}~\bibnamefont{Farrow}},
  \bibinfo{author}{\bibfnamefont{S.}~\bibnamefont{Diaz-Moreno}},
  \bibnamefont{and} \bibinfo{author}{\bibfnamefont{C.}~\bibnamefont{Ponchut}},
  \bibinfo{journal}{Nucl. Instrum. Methods} \textbf{\bibinfo{volume}{551}},
  \bibinfo{pages}{27} (\bibinfo{year}{2005}).

\bibitem[{\citenamefont{Visser et~al.}(2001)\citenamefont{Visser,
  Anxolabehere-Mallart, Bergmann, Glatzel, Robblee, Cramer, Girerd, Sauer,
  Klein, and Yachandra}}]{Hend2001}
\bibinfo{author}{\bibfnamefont{H.}~\bibnamefont{Visser}},
  \bibinfo{author}{\bibfnamefont{E.}~\bibnamefont{Anxolabehere-Mallart}},
  \bibinfo{author}{\bibfnamefont{U.}~\bibnamefont{Bergmann}},
  \bibinfo{author}{\bibfnamefont{P.}~\bibnamefont{Glatzel}},
  \bibinfo{author}{\bibfnamefont{J.~H.} \bibnamefont{Robblee}},
  \bibinfo{author}{\bibfnamefont{S.~P.} \bibnamefont{Cramer}},
  \bibinfo{author}{\bibfnamefont{J.-J.} \bibnamefont{Girerd}},
  \bibinfo{author}{\bibfnamefont{K.}~\bibnamefont{Sauer}},
  \bibinfo{author}{\bibfnamefont{M.~P.} \bibnamefont{Klein}}, \bibnamefont{and}
  \bibinfo{author}{\bibfnamefont{V.~K.} \bibnamefont{Yachandra}},
  \bibinfo{journal}{J. Am. Chem. Soc.} \textbf{\bibinfo{volume}{123}},
  \bibinfo{pages}{7031} (\bibinfo{year}{2001}).

\bibitem[{\citenamefont{Bousseksou et~al.}(2002)\citenamefont{Bousseksou,
  Boukheddaden, Goiran, Consejo, Boillot, and Tuchagues}}]{Bousseksou2002}
\bibinfo{author}{\bibfnamefont{A.}~\bibnamefont{Bousseksou}},
  \bibinfo{author}{\bibfnamefont{K.}~\bibnamefont{Boukheddaden}},
  \bibinfo{author}{\bibfnamefont{M.}~\bibnamefont{Goiran}},
  \bibinfo{author}{\bibfnamefont{C.}~\bibnamefont{Consejo}},
  \bibinfo{author}{\bibfnamefont{M.-L.} \bibnamefont{Boillot}},
  \bibnamefont{and} \bibinfo{author}{\bibfnamefont{J.-P.}
  \bibnamefont{Tuchagues}}, \bibinfo{journal}{Phys. Rev. B}
  \textbf{\bibinfo{volume}{65}}, \bibinfo{pages}{172412}
  (\bibinfo{year}{2002}).

\bibitem[{\citenamefont{Bousseksou et~al.}(2004)\citenamefont{Bousseksou,
  Varret, Goiran, Boukheddaden, and Tuchagues}}]{Bousseksou2004}
\bibinfo{author}{\bibfnamefont{A.}~\bibnamefont{Bousseksou}},
  \bibinfo{author}{\bibfnamefont{F.}~\bibnamefont{Varret}},
  \bibinfo{author}{\bibfnamefont{M.}~\bibnamefont{Goiran}},
  \bibinfo{author}{\bibfnamefont{K.}~\bibnamefont{Boukheddaden}},
  \bibnamefont{and}
  \bibinfo{author}{\bibfnamefont{J.}~\bibnamefont{Tuchagues}},
  \emph{\bibinfo{title}{Spin Crossover in Transition Metal Compounds II}}
  (\bibinfo{publisher}{Springer Berlin / Heidelberg}, \bibinfo{year}{2004}),
  chap. \bibinfo{chapter}{The Spin Crossover Phenomenon Under High Magnetic
  Field}.

\bibitem[{\citenamefont{Bousseksou et~al.}(2000)\citenamefont{Bousseksou,
  Negre, Goiran, Salmon, Tuchagues, Boillot, Boukheddaden, and
  Varret}}]{Bousseksou2000}
\bibinfo{author}{\bibfnamefont{A.}~\bibnamefont{Bousseksou}},
  \bibinfo{author}{\bibfnamefont{N.}~\bibnamefont{Negre}},
  \bibinfo{author}{\bibfnamefont{M.}~\bibnamefont{Goiran}},
  \bibinfo{author}{\bibfnamefont{L.}~\bibnamefont{Salmon}},
  \bibinfo{author}{\bibfnamefont{J.-P.} \bibnamefont{Tuchagues}},
  \bibinfo{author}{\bibfnamefont{M.-L.} \bibnamefont{Boillot}},
  \bibinfo{author}{\bibfnamefont{K.}~\bibnamefont{Boukheddaden}},
  \bibnamefont{and} \bibinfo{author}{\bibfnamefont{F.}~\bibnamefont{Varret}},
  \bibinfo{journal}{Eur. Phys. J. B} \textbf{\bibinfo{volume}{13}},
  \bibinfo{pages}{451} (\bibinfo{year}{2000}).

\bibitem[{\citenamefont{Negre et~al.}(2000)\citenamefont{Negre, Goiran,
  Bousseksou, Haasnoot, Boukheddaden, Askenazy, and Varret}}]{Negre2000}
\bibinfo{author}{\bibfnamefont{N.}~\bibnamefont{Negre}},
  \bibinfo{author}{\bibfnamefont{M.}~\bibnamefont{Goiran}},
  \bibinfo{author}{\bibfnamefont{A.}~\bibnamefont{Bousseksou}},
  \bibinfo{author}{\bibfnamefont{J.}~\bibnamefont{Haasnoot}},
  \bibinfo{author}{\bibfnamefont{K.}~\bibnamefont{Boukheddaden}},
  \bibinfo{author}{\bibfnamefont{S.}~\bibnamefont{Askenazy}}, \bibnamefont{and}
  \bibinfo{author}{\bibfnamefont{F.}~\bibnamefont{Varret}},
  \bibinfo{journal}{Synthetic Metals} \textbf{\bibinfo{volume}{115}},
  \bibinfo{pages}{289 } (\bibinfo{year}{2000}).

\bibitem[{\citenamefont{Hauser}(2004)}]{Hauser2004}
\bibinfo{author}{\bibfnamefont{A.}~\bibnamefont{Hauser}},
  \emph{\bibinfo{title}{Spin Crossover in Transition Metal Compounds II}}
  (\bibinfo{publisher}{Springer Berlin / Heidelberg}, \bibinfo{year}{2004}),
  chap. \bibinfo{chapter}{Light-Induced Spin Crossover and the High-Spin
  Low-Spin Relaxation}.

\end{thebibliography}

\end{document}